\documentclass[aps,prb,reprint,floatfix,groupedaddress,superscriptaddress,showpacs]{revtex4-1}

\usepackage{lipsum}
\usepackage{graphicx}
\usepackage{natbib}
\usepackage[utf8]{inputenc}
\usepackage{tikz}
\usepackage{amsmath}
\usepackage{empheq}
\usepackage{geometry}
\usepackage{amssymb}

\newcommand{\srni}{Sr$_3$NiIrO$_6$}
\newcommand{\am}{\AA$^{-1}$}
\newcommand{\nic}{Ni$^{2+}$}
\newcommand{\iri}{Ir$^{4+}$}
\newcommand{\vect}[1]{{\bf{#1}}}

\begin{document}

\title{Frustrated Ising chains on the triangular lattice in \srni}

\author{S. Toth}
\email{sandor.toth@psi.ch}
\affiliation{Laboratory for Neutron Scattering and Imaging, Paul Scherrer Institut (PSI), CH-5232 Villigen, Switzerland}

\author{W. Wu}
\affiliation{Department of Physics and Astronomy and London Centre for Nanotechnology, University College London, Gower Street, London WC1E 6BT, UK}

\author{D. T. Adroja}
\email{devashibhai.adroja@stfc.ac.uk}
\affiliation{ISIS Facility, STFC, Rutherford Appleton Laboratory, Chilton, Oxfordshire OX11 0QX, UK}
\affiliation{Highly Correlated Matter Research Group, Physics Department, University of Johannesburg, PO Box 524, Auckland Park 2006, South Africa}

\author{S. Rayaprol}
\affiliation{UGC-DAE Consortium for Scientific Research, Mumbai Center, R-5 Shed, BARC, Trombay, IN-400085 Mumbai, India}

\author{E. V. Sampathkumaran}
\affiliation{Tata Institute of Fundamental Research, Homi Bhabha Road, Colaba, Mumbai 400005, India}

\date{\textrm{\today}}


\begin{abstract}
Inelastic neutron scattering study on the spin-chain compound \srni ~reveals gapped quasi-1D magnetic excitations. The observed one-magnon band between 29.5 and 39 meV consists of magnon modes of the \nic ions. The fitting of the spin wave spectrum reveals strongly coupled Ising-like chains along the $c$-axis that are weakly coupled into a frustrated triangular lattice in the $ab$-plane. The magnetic excitations survive up to 200 K well above the magnetic ordering temperature of $T_N \sim 75$ K, also indicating a quasi-1D nature of the magnetic interactions in \srni. Our microscopic model is in good agreement with ab initio electronic structure calculations and explains the giant spin flip field observed in bulk magnetization measurements.
\end{abstract}

\maketitle

\section{Introduction}

Low-dimensional and geometrically frustrated spin systems exhibit some of the most interesting physical phenomena seen in condensed matter physics. Due to the low site connectivity and competing interactions classical order is often suppressed by quantum and thermal fluctuations giving rise to novel ground states and quasiparticle excitations. Beside the spin liquid states \cite{Balents2010} where no long range order exists, certain geometries, such as the Ising model on the stacked triangular lattice antiferromagnet (TLA), possess partially disordered ground states \cite{NihatBerker1984,Jiang2006}. Theory predicts two phases beside the paramagnetic phase. The first phase consists of two antiferromagnetically ordered sublattices and a disordered third one, while the low temperature phase has one fully ordered site and two partially ordered site with opposite moment direction and zero net moment. The most prominent experimental realizations of the Ising model on the TLA are CsCoBr$_3$ \cite{Yelon1975,Farkas1991,Mao2002} and Ca$_3$Co$_2$O$_6$ \cite{Agrestini2008,Jain2013}. Both compounds have strongly coupled Ising chains perpendicular to the triangular plane. While CsCoBr$_3$ has antiferromagnetic chains, Ca$_3$Co$_2$O$_6$ has ferromagnetic chains producing magnetization plateaus \cite{Kageyama1997,Zhao2010,Takubo2005a,Chapon2009,Sampathkumaran2004a,Sampathkumaran2004}. We propose a novel frustrated system with Ising spins on the stacked TLA with strongly coupled ferrimagnetic chains of alternating \nic\ and \iri ions: \srni\cite{Flahaut2003a,Sarkar2010,Mikhailova2012,Ou2014,Lefrancois2014a}.

\begin{figure}[!htb]
    \centering
	\includegraphics[width = \linewidth]{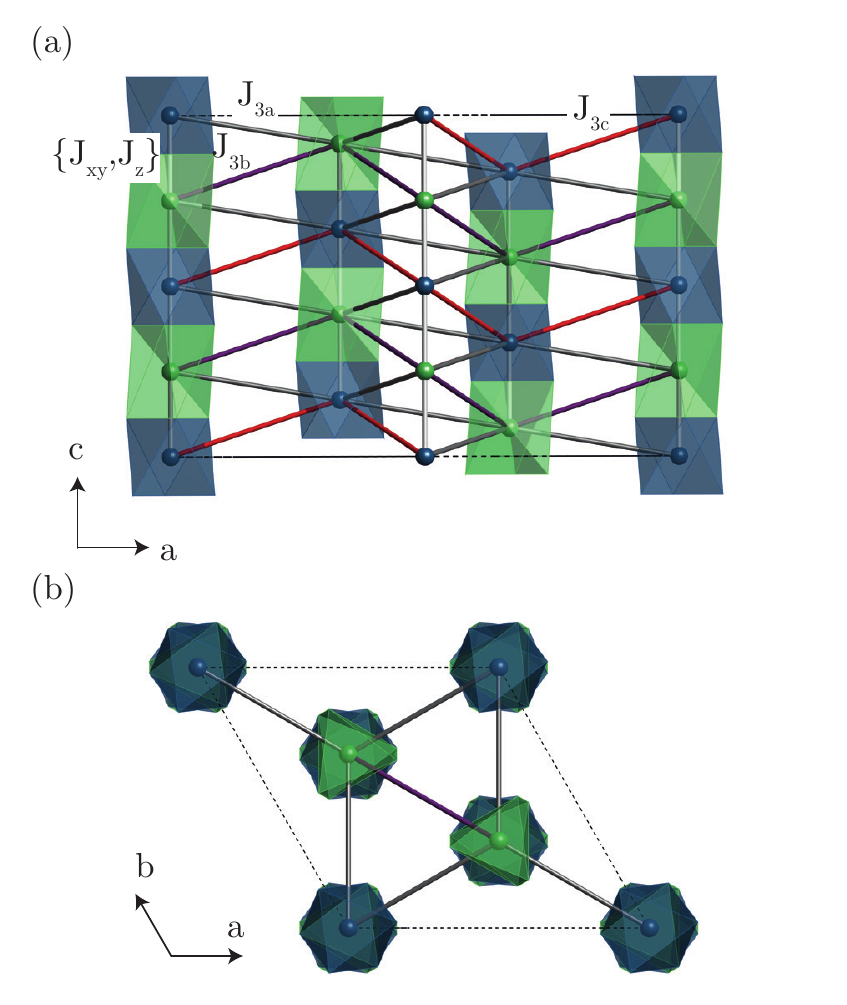}
	\caption{(Color online) Crystal structure of \srni, showing the NiO$_6$ trigonal prism (light green) and IrO$_6$ octahedra (dark green). The white vertical bonds are the first neighbor anisotropic exchange interactions $J_{xy}$ and $J_z$, the zig-zag bonds with three different colors denote the three inequivalent interchain coupling $J_{3a}$, $J_{3b}$ and $J_{3c}$.}
	\label{fig:struct}
\end{figure}

\srni\ together with Ca$_3$Co$_2$O$_6$ belong to a larger family of spin-chain systems with general formula A$_3$MM'O$_6$ ($A=$ alkaline-earth metal, $M/M'=$ transition metals) that have attracted much attention in recent years, due to their reduced dimensionality.  Sr$_3$ZnRhO$_6$ \cite{Hillier2011}, Sr$_3$CuIrO$_6$ \cite{Yin2013}, Ca$_3$CoRhO$_6$ \cite{Sampathkumaran2002}, Sr$_3$CuRhO$_6$ \cite{Sampathkumaran2007}, Ca$_3$CoRhO$_6$ \cite{Niitaka2001a,Basu2014} are the most studied ones showing magnetization jumps, large thermoelectric power and magnetoresistance \cite{Raquet2002,Maignan2003}. The crystal structure consists of 1D chains that are oriented along the $c$-axis and arranged in a triangular lattice in the $ab$ plane, see Fig.\ \ref{fig:struct}. The chains are formed by alternating face-sharing MO$_6$ trigonal prism and M’O$_6$ octahedra and intercalated by A$^{2+}$ cation, thus forming a triangular arrangement.

Beside the strongly one dimensional crystal structure these compounds possess strong spin anisotropy. It originates from either the single ion property of the MO$_6$ site as for Ca$_3$Co$_2$O$_6$ where the weak spin orbit coupling (SOC) can induce a large orbital moment on the high spin carrier Co$^{3+}$ due to the distorted symmetry of the trigonal prism \cite{Wu2005}. Beside for heavier transition metals such as rhodium or iridium, the strong SOC can induce anisotropic exchange interaction as for Sr$_3$CuIrO$_6$ \cite{Yin2013}. In \srni\ both of these mechanisms are potentially active, where the \nic\ occupy the trigonal prism site with $d^8$ electronic configuration ($S=1$) and the \iri\ taking the octahedral site with a novel $J_{eff}=1/2$ electronic state. Recent ab initio results have shown that the coupling along the chain is AFM \cite{Ou2014} but only if the SOC is taken into account which results in a ferrimagnetic order due to the different moment sizes of the two magnetic ions.

The observed low temperature magnetic structure of \srni\ is also intriguing. Lefrançois et al.\ found a $\vect{k}=(0,0,1)$ magnetic order with strongly reduced magnetic moments using neutron diffraction \cite{Lefrancois2014a}. The refinement revealed that all magnetic moments are parallel to the $c$-axis and within each Ni-Ir chain the moments are ferrimagnetically ordered. However from diffraction alone the global phase of the structure cannot be determined. This gives two qualitatively different solutions (with a continuum of possibilities between). In the first solution one chain in the unit cell is fully ordered, while the other two has ordered moments reduced by half and the ferrimagnetic moment pointing in the opposite direction. The second solution has two fully ordered chains with opposite ferrimagnetic moment and a completely disordered third chain. Both of these structures are predicted theoretically for the Ising model on the stacked TLA.

Previous RIXS study on \srni\ found a band of magnetic excitations centered at 95 meV\cite{Lefrancois2015a} at 10 K. The observed inelastic intensity and the size of the gap was gradually decreasing with increasing temperature. At room temperature the excitations were centered at 50(5) meV. Since the excitations were measured using the resonant $L_3$-edge of iridium, the experiment shows selectively the magnetic signal only on the iridium atoms. The authors did not discuss, whether the width of the observed excitations is resolution limited.

A good understanding of the magnetic properties of \srni\ requires both high energy-resolution probing technique and spin wave calculations, in which the SOC is presented in the form of anisotropic exchange interaction and spin anisotropies. In the present work, we show a combination of inelastic neutron scattering (INS) measurements and spin wave calculation for \srni\ revealing a very anisotropic exchange Hamiltonian and an effective Ising  model on the stacked triangular lattice.

\section{Experimental details}

Polycrystalline sample of \srni\ was prepared by solid-state reactions of NiO, IrO$_2$ and SrCO$_3$. The \srni\ sample used in the present study is the same sample used in our previous neutron diffraction study \cite{Lefrancois2014a}. The X-ray powder diffraction study at 300 K and neutron diffraction study at 100 K show that the \srni\ sample was single phase and crystallized in the space group R$\overline{3}$c. The INS measurements were performed on 6 g of sample using the high count rate time-of-flight chopper spectrometer, MERLIN at the ISIS facility, UK. To reduce the neutron absorption problem from iridium, we filled the fine powder of \srni\ in an aluminum foil envelope rolled into a cylindrical shape with a diameter of 40 mm (and a height of 45 mm) and then inserted into a cylindrical aluminum can and finally mounted into a closed-cycle refrigerator under He-exchange gas. The average sample thickness was less than 1 mm. We corrected the data for neutron absorption, which was calculated to be 15\% of the incident beam. The INS measurements were carried out with various incident neutrons energies: $E_i=15$, 80, 150 and 500 meV and temperatures between 5 K and 300 K. We also measured a standard vanadium sample at the same set of incident energies to determine the energy resolution at the elastic line and to convert the intensities into normalized units of cross section, mbr/sr/meV/f.u., where f.u. stands for formula unit of \srni.

\section{Results}

\begin{figure}[!htb]
    \centering
	\includegraphics[width = \linewidth]{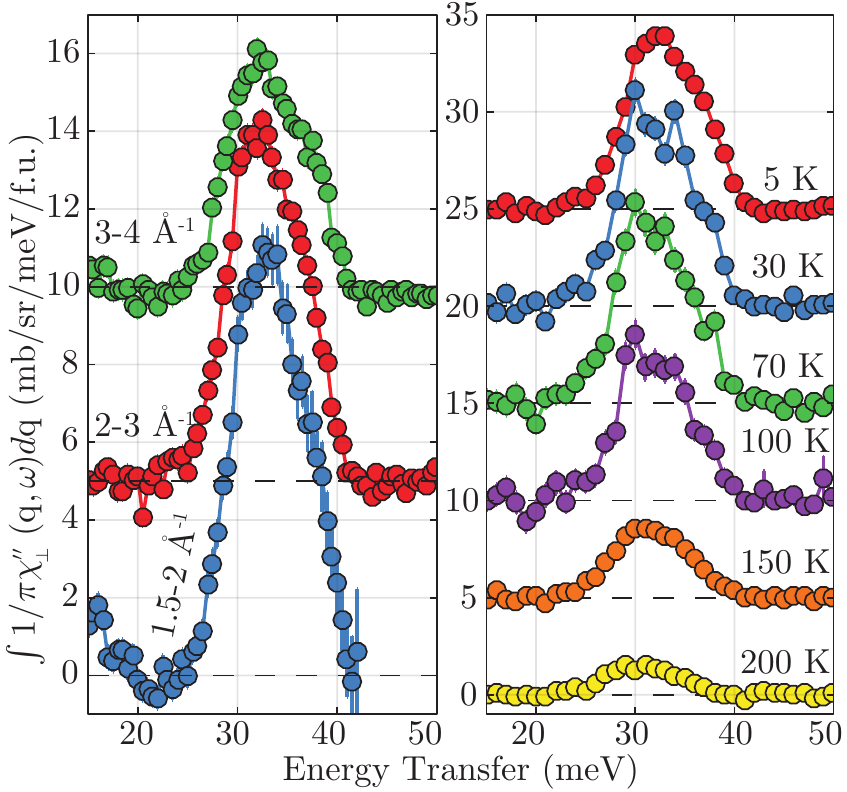}
	\caption{(Color online) Imaginary part of the magnetic dynamical susceptibility of \srni\ measured using inelastic neutron scattering on polycrystalline sample collected with incident neutron energy of 80 meV and after the subtraction of the non-magnetic background. The signal is in absolute units. (a) Data cuts measured at 5 K and integrated for different $Q$ ranges, (b) data cuts measured at different temperatures and integrated between 2 and 3 \am. Note that the slightly negative signal is an artifact of the background subtraction.}
	\label{fig:peak}
\end{figure}

\begin{figure}[!htb]
    \centering
	\includegraphics[width = \linewidth]{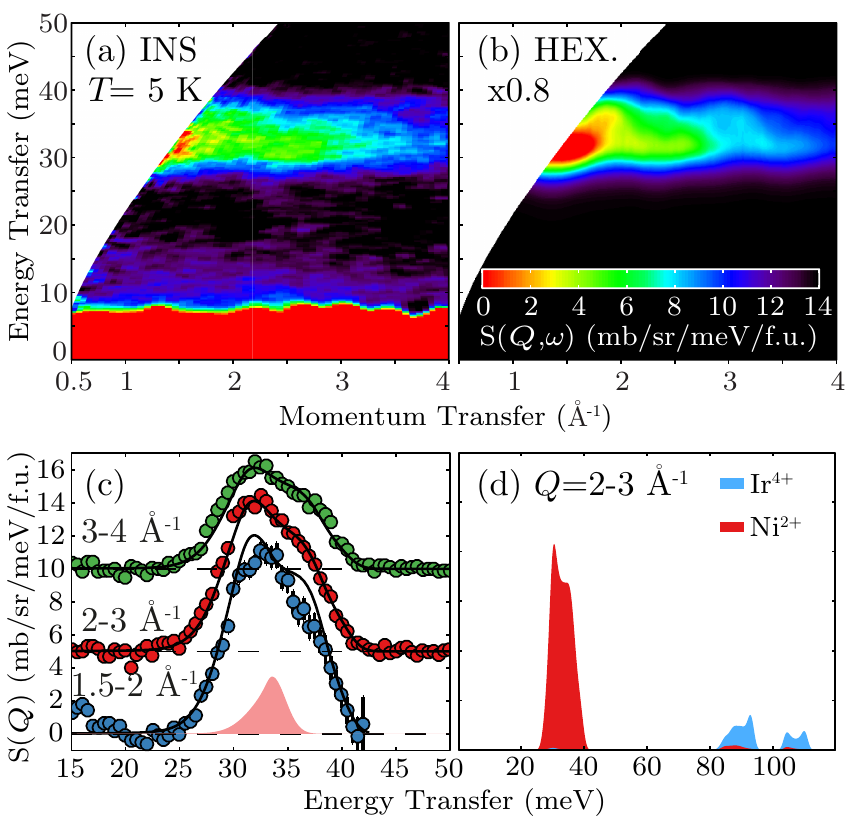}
		\caption{(Color online)  (a) The magnetic scattering of \srni\ at 5 K obtained after subtracting phonon scattering, strong scattering below 10 meV is due to the incoherent background. (b) The simulated spin wave scattering at 5 K using SpinW program with the parameters of the best fitting hexagonal structure, intensity scaled with a factor of 0.8 to fit the data, the solid peak at the bottom shows the instrumental energy resolution at 32.5 meV. (c) Cuts at different $Q$ ranges of both the data (color circles) and simulation (black lines). (d) The complete spin wave spectrum, blue and red areas show the intensity of spin waves localized on iridium and nickel ions respectively.}
	\label{fig:fit}
\end{figure}

\begin{figure}[!htb]
    \centering
	\includegraphics[width = \linewidth]{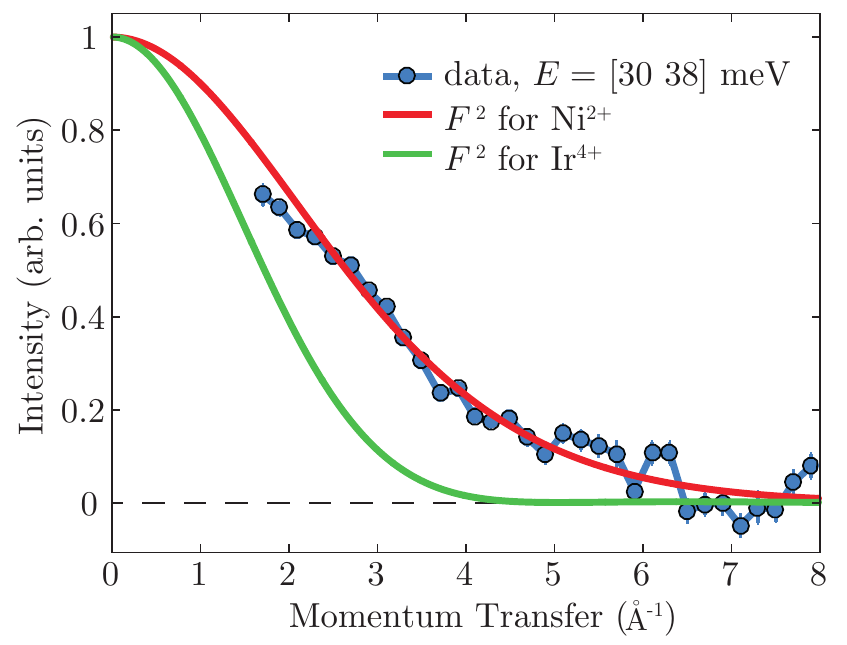}
	\caption{(Color online) Momentum transfer dependence of the integrated inelastic magnetic signal denoted by blue dots. Red and green lines denote the squared  $\langle j_0\rangle$ magnetic form factor of \nic\ and \iri\ ions\cite{Kobayashi2011}.}
	\label{fig:f2}
\end{figure}

The inelastic neutron scattering data reveals two types of excitations. At momentum transfers above 4 \am\ strong inelastic scattering was observed with intensity increasing proportionally to the momentum transfer square ($|Q|^2$) and with increasing temperature. These properties clearly indicate scattering due to phonons. We also observed inelastic scattering below 4 \am\ between 29.5 and 39 meV that decreases in intensity with increasing momentum transfer and temperature, see Fig.\ \ref{fig:peak}. We assign these excitations to magnetic scattering. To separate the magnetic and phonon scattering we collected data at room temperature that contains only phonon scattering and used this data to subtract the phonons from the magnetic signal (for details see Appendix \ref{sec:fit}). The magnetic signal is well defined in the magnetically ordered phase below $T_N=75$ K and survives up to 200 K with gradually decreasing intensity, see Fig.\ \ref{fig:peak}(b). At low temperatures the signal is due to spin wave scattering, while in the paramagnetic phase is due to low dimensional scattering of the strongly correlated chains. The momentum transfer dependence of the magnetic signal follows the $\langle j_0 \rangle$ (spin only) form factor of the \nic\ ions, see Fig.\ \ref{fig:f2}. The energy width of the excitations far exceeds the instrumental resolution which is a sign of dispersive modes. Moreover the peak is asymmetric with a well developed shoulder at the high energy side.

We also measured inelastic neutron scattering using 150 meV and 500 meV incident energy to confirm the previous RIXS results, however we only found a very weak scattering centered at 87 meV. Due to the low signal to noise ratio we could not unambiguously assign this scattering to the magnetism of the iridium ions. There are two reasons why the iridium signal is so weak compared to the nickel signal for INS. Due to the smaller spin-1/2 effective quantum number of \iri\ it gives half the intensity compared to the spin-1 \nic\ ions. Beside at increasing energies, the lowest momentum transfer that is measurable by a direct time of flight instrument is also increasing which gives strongly reduced intensity due to the magnetic form factor. In the following we assume the upper iridium mode is centered at 95 meV with unknown bandwidth.

\section{Analysis}

The observed magnetic excitations of \srni\ in the ordered phase can be modeled using linear spin wave theory. We will do this in two steps. First we propose a one dimensional Ni-Ir alternating chain model, where we neglect the interchain couplings. Afterwards to improve the model we will introduce additional magnetic exchange interactions between the chains and we will show that the interchain interactions are necessary to adequately fit the data.

To be able to model the magnetic excitations using linear spin wave theory, we need a classical magnetic ground state. However experimentally determined magnetic structures are incompatible with a zero temperature classical ground state. In the following we propose model Hamiltonians with ground state close to the observed one and we will show that the calculated excitation spectrum is insensitive to the details of the magnetic ground state.

The simplest model Hamiltonian to describe the observed spin waves is the Ni-Ir alternating chain along the $c$-axis. Due to the 3-fold symmetry along the $c$-axis, the most general spin Hamiltonian (up to two spin exchanges) allowed by the symmetry is the following:

\begin{align}
\label{eq:eq1}
\mathcal{H} =& \displaystyle\sum_{i} J_{xy}\left(S_i^xS_{i+1}^x+S_i^yS_{i+1}^y\right)  +J_zS_i^zS_{i+1}^z\\
&+ \sum_{i=2k} A S_i^zS_i^z\nonumber + \sum_i D_i \left( S_i^xS_{i+1}^y-S_i^yS_{i+1}^x\right),
\end{align}

where $S_i$ denotes the $J_{eff}=1/2$ quantum number of iridium ions if $i=2k+1$ and the $S=1$ spin of nickel ions if $i=2k$. Also we assumed that only the nickel ions have single ion anisotropy ($A$). The Dzyaloshinskii-Moriya (DM) interaction is also allowed with the DM vector parallel to the $c$-axis $\vect{D}=(0,0,D_i)$ and the sign of $D_i$ is positive for $i\in\{4k+1,4k+2\}$ and negative otherwise.

The classical zero temperature ground state of the above Hamiltonian for dominating antiferromagnetic $J_z$ exchange is the ferrimagnetic chain in agreement with neutron diffraction experiments. If the exchange interactions are Heisenberg type ($J_{xy}=J_z$) as one would expect for transition metals the excitation spectrum would have a zero gap (up to small value due to negative $A$). However we observed a spin wave gap much larger than the bandwidth of the excitations which implies that $J_z>J_{xy}$. This anisotropic exchange is also compatible with the strong spin orbit coupling expected for iridium. The energy width of the measured magnon band is due to a dispersive mode, which is related to the size of $J_{xy}$ and the double peak structure is due to the Van Hove singularities at the bottom and top of the magnon band smeared by the finite instrumental resolution.

The separation of the nickel and iridium spin wave modes is the consequence of the different Weiss field of the magnetic ions that happens even for completely isotropic interactions. Since the Weiss field is linear with the spin quantum number of the neighbors, it is larger on the iridium ions than on the nickel ions. Thus the spin wave modes that mainly localized on the Ir acquire a larger gap than on the Ni. Due to the large energy separation the mixing of the spin wave modes is negligible. This also means that we cannot fit any coupling between iridium spins ($J_{2a}$ and $J_{3c}$) since it only influence the upper iridium spin wave band of which we know only the position but not the shape as a function of energy.

In order to fit the observed powder data, we extracted a single cut through the inelastic signal integrated from 2 \AA$^{-1}$ to 3 \AA$^{-1}$ and binned in energy with 0.5 meV steps. This data was compared with the calculated spin-spin correlation function of the above Hamiltonian using linear spin wave theory with SpinW \cite{Toth2014a}. The powder averaged neutron scattering cross section is calculated using the equation:
\begin{align}
	I(Q,\omega) = \frac{1}{4\pi Q^2} \int_{|\vect{q}|=Q}\sum_{\alpha,\beta}(1-\hat{q}^\alpha\hat{q}^\beta) S^{\alpha\beta}(\vect{q},\omega)d\vect{q},
\end{align}
where $S^{\alpha\beta}(\vect{q},\omega)$ is the spin-spin correlation function including the magnetic form factors of the different ions and the integration runs in reciprocal space covering the sphere with radius $Q$. In the simulation we numerically integrated over 987 $\vect{q}$-points covering the $Q=2.5$ \am\ radius sphere with near uniform spacing (the points were generated according to \cite{Hannay2004}). We chose to use a fixed set of $Q$ points, since a random distribution of $Q$ points would make the fit unstable. The calculated powder averaged spectrum was convoluted with the instrumental energy resolution function (see Appendix \ref{sec:res}). This calculated data was then fitted to the measured data using weighted least squares refinement. To determine the optimal parameters of Eq.\ \ref{eq:eq1}, we applied a stochastic optimization method (described in Appendix \ref{sec:fit}).

\begin{ruledtabular}
\begin{table}[!htb]
	\centering
	\caption{Comparison of the best fitting parameters of different spin wave models of \srni.}
	\label{tab:fitres}
		\begin{tabular}{l|c|c|c|c}
			& chain & hexagonal & stripy & FM \\
			\hline
			$J_{xy}$ (meV)   & 22.7  & 21.6  & 11.7  & 15.6   \\
			$J_z$    (meV)   & 49.8  & 46.6  & 45.3  & 42.6   \\
			$A$      (meV)   & 6.31  & 4.95  & 7.19  & 5.17   \\
			$J_{2b}$ (meV)   & -     & -     & -1.50 & -0.842 \\			
			$J_{3a}$ (meV)   & -     & -2.83 & -2.78 & 2.02   \\
			$J_{3b}$ (meV)   & -     & -1.37 & -1.05 & 0.872  \\
			$J_{tri}$ (meV)  & 0     & 1.46  & 1.47  & -1.15  \\
			\hline
			$\chi_{red}^2$   & 14.15 & 1.53  & 1.86  & 2.91   \\
		\end{tabular}
\end{table}
\end{ruledtabular}

The parameters of the best fitting single chain model is shown in the first column of Tab.\ \ref{tab:fitres}. The best fit is achieved when the Dzyaloshinskii-Moriya interactions were constrained to zero. We also included an isotropic second neighbor interactions along the chains $J_{2a}$ and $J_{2b}$, however it did not improve the $\chi_{red}^2$ value. Although this model reproduced the main features of the spectrum, the large $\chi_{red}^2$ value reveals that the model has to be improved. It is important to note that since we only have a poorly resolved upper iridium band, we cannot fit iridium--iridium couplings that will only influence the shape of the upper band.

\begin{ruledtabular}
\begin{table}[!htb]
	\centering
	\caption{List of symmetry allowed exchange couplings in \srni. $r$ denotes the bond length at 100 K\cite{Lefrancois2014a}, $n$ denotes the number of bond per unit cell, $J_S$ denotes the symmetric part of the exchange matrix (diagonal elements are given, otherwise no symmetry constraint) and $J_A$ denotes the antisymmetric part of the exchange matrix (given as a DM vector).}
	\label{tab:bond}
		\begin{tabular}{l|l|r|r|c|c}
			label     & atoms  & $r$ (\AA) & $n$& $J_{S}$     			 & $J_{A}$ 	\\
			\hline
			$J_1$    & Ni--Ir 	& 2.791     & 12 & $(J_{xy},J_{xy},J_z)$ & $(0,0,D_1)$  \\
			$J_{2a}$ & Ir--Ir 	& 5.583     &  6 & $(a,a,b)$    		 & $(0,0,c)$    \\
			$J_{2b}$ & Ni--Ni 	& 5.583     &  6 & $(a,a,b)$    		 & $(0,0,0)$    \\
			$J_{3a}$ & Ni--Ir 	& 5.626     & 36 & general  			 & $(a,b,c)$    \\
			$J_{3b}$ & Ni--Ni 	& 5.852     & 18 & general  			 & $(0,0,0)$    \\
			$J_{3c}$ & Ir--Ir 	& 5.852     & 18 & general   			 & $(a,b,c)$    \\
		\end{tabular}
\end{table}
\end{ruledtabular}

To improve the model, we have to take into account further neighbor interactions that couple the chains. The shortest interchain interactions couple the chains into a frustrated triangular lattice. There are three bonds with similar length denoted by $J_{3a}$, $J_{3b}$ and $J_{3c}$, see Tab.\ \ref{tab:bond}. Due to the larger length, we expect that these couplings are much weaker than the ones along the chain, thus we simply model them as being Heisenberg type (isotropic). Altogether there are 72 interchain bonds per unit cell.

\begin{figure}[!htb]
    \centering
	\includegraphics[width = \linewidth]{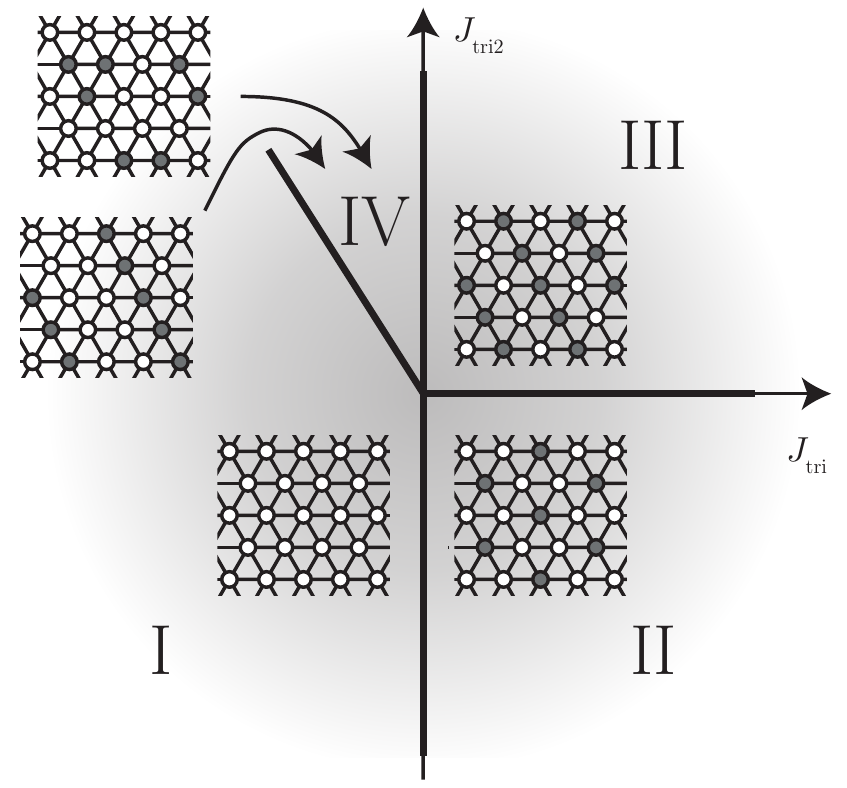}
	\caption{Phase diagram of the triangular lattice Ising model with interactions up to the second neighbor. $J_{tri}$ is the effective coupling between the Ni-Ir chains (see text) and $J_{tri2}$ is the effective coupling between second neighbor chains. Grey and white circles denote spins with opposite directions.}
	\label{fig:phase}
\end{figure}

To determine the possible classical ground states of the coupled chain model, we introduce an effective Ising model. Since the anisotropy along the chains is very strong, in the classical ground state the degree of freedom of a chain is equivalent to that of a single Ising spin. If we group the neighboring iridium and nickel spins together along the chain, we get an effective ferromagnetic spin-1/2 chain. Also the coupling between the chains can be mapped to couplings between the Ising spins. If we index the three chains in the unit cell with 1 for position $(0,0,z)$, 2 for $(2/3,1/3,z)$ and 3 for $(1/3,2/3,z)$ and define the sign of the first nickel magnetic moment along $z$-axis as $\varphi_i$ on the $i$th chain, then the classical energy per formula unit for $k=0$ magnetic structures is:
\begin{align}
	E &= \left(\varphi_1\varphi_2+\varphi_1\varphi_3+\varphi_2\varphi_3\right)J_{tri},\\\nonumber
	J_{tri} &=J_{3b}S_{Ni}^2+J_{3c}S_{Ir}^2-2J_{3a}S_{Ni}S_{Ir},
\end{align}
where $J_{tri}$ is an effective coupling between the ferromagnetic Ising chains creating a triangular lattice perpendicular to the chains. The possible classical ground states of the Ising model on the triangular lattice is well known. If $J_{tri}$ is ferromagnetic, the ground state is a simple ferromagnet, while for antiferromagnetic $J_{tri}$ the system is frustrated with disordered ground state \cite{Wannier1950}. However if we introduce a vanishingly small effective coupling between second neighbor chains, we would get two types of ordered phase, see Fig.\ \ref{fig:phase} according to \cite{Tanaka1975}: hexagonal (phase-II) and stripy (phase-III). It is important to note that the stripy phase would give a non-zero $k$ magnetic structure in respect to the crystallographic unit cell which disagrees with the observed $k=0$ inplane ordering wave vector. However we will keep both models in order to see how sensitive is the result to the type of ground state. We can also calculate the magnetic moment per formula unit along the $z$-axis for the $k=0$ structures:
\begin{align}
	M_z = -\frac{1}{3}\left(\varphi_1+\varphi_2+\varphi_3\right)\left(M_{Ni}-M_{Ir}\right),
\end{align}
where $M_{Ni}$ and $M_{Ir}$ are the atomic magnetic moment of nickel and iridium respectively. Assuming the gyromagnetic ratio $g=2$ for both ions, the magnetic moment per formula unit is $1\mu_B$, $0.3\mu_B$ and $0$  for the ferromagnetic, hexagonal and stripy structures respectively. The magnetization value of the hexagonal ordering agrees well with the experimental field-cooled magnetization value of $0.25\mu_B$ \cite{Lefrancois2014a}.

\begin{figure}[!htb]
    \centering
	\includegraphics[width = \linewidth]{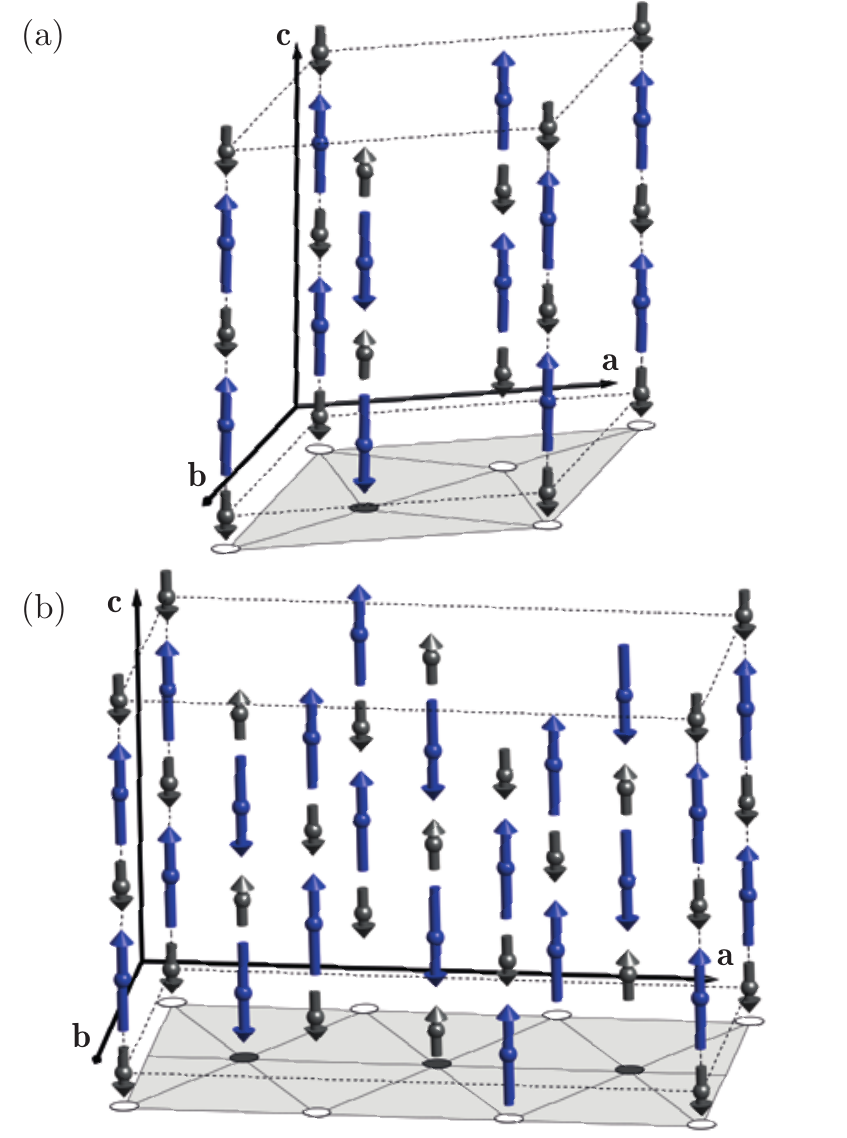}
	\caption{(Color online) Magnetic structures used for the modeling of the (a) hexagonal structure and (b) stripy structure. The underlying gray parallelepipeds show the equivalent Ising model with black and white spheres for the down and up spins.}
	\label{fig:mag}
\end{figure}

To calculate the spin wave spectrum for the three different ground states we have to return to the original lattice of \srni\ since the equivalence between the effective Ising model and \srni\ holds strictly only for the classical ground state. For the hexagonal and ferromagnetic structures the crystallographic unit cell is the minimal cell that can describe the ground state, while for the stripy structure the smallest cell is rectangular in the $ab$-plane with lattice vectors of (1,0,0) and (1,2,0) in units of the crystallographic lattice vectors. The two frustrated magnetic structures are plotted in Fig.\ \ref{fig:mag}.

The best fits of both the stripy and hexagonal structures with interchain interactions are significantly better than the single chain model, see Tab.\ \ref{tab:fitres}. We could achieve good fits with both ground states, although the ideal fit should give $\chi^2_{red}\approx1$. The increased $\chi^2_{red}$ values can be attributed to systematic errors, such as the non-ideal background subtraction and the non-exact definition of the energy resolution function. The fits reveal that the effective exchange interaction $J_{tri}$ between the chain is antiferromagnetic and the fit value is not sensitive to the ground state magnetic structure.

\section{Discussion}

The observed magnetic excitations of \srni\ survive up to 200 K which is a sign of low dimensionality. Our linear spin wave model indeed revealed strong coupling along the chains with strongly anisotropic exchange matrix. Beside the magnetic peak as a function of energy is much broader than the instrumental resolution, which is the sign of dispersive spin wave modes along and between the chains. The fitting of the coupled chain model parameters to the experimental data revealed essential information regarding the magnetism of \srni. Considering the leading terms in the Hamiltonian, our fit results provide a reliable answer. The largest term in the Hamiltonian is the first neighbor antiferromagnetic exchange interaction between nickel and iridium ions along the $c$-axis in agreement with other experiments \cite{Flahaut2003a,Lefrancois2014a} and theory \cite{Zhang2010,Ou2014}. Moreover we found that the exchange interaction is strongly anisotropic which was not shown before and which is compatible with the strong spin-orbit coupling of the iridium. The exchange values with a conservative error estimation are $J_{xy}=21.6(10)$ meV and $J_z=46.6(13)$ meV. We also found that the nickel ion shows easy plane anisotropy in the $ab$-plane, with a value of $A=5.0(10)$ meV, while anisotropy on the iridium site is much smaller. Assuming single ion anisotropy on the iridium site alone cannot describe the data. These values show the strong uniaxial magnetism of \srni\ originates from the anisotropic exchange interaction between iridium and nickel. The unusually strong anisotropy of the nickel ion is the result of the strongly distorted local environment within the strained trigonal prism of oxygens. A similar value with opposite sign ($D+J_z=-7.20(2)$ meV) was found in Ca$_2$Co$_2$O$_6$ by inelastic neutron scattering and ab initio calculations \cite{Wu2005,Jain2013}. Additional terms in the single chain Hamiltonian do not play an important role, regarding the spin wave excitations. Dzyaloshinskii-Moriya and further neighbor interactions along the chain can be also neglected with a good approximation. The strongly anisotropic ferrimagnetic single chain model allows us to rewrite it into an equivalent Ising model, where each chain would be equivalent to a large Ising spin with $M\approx1\mu_B$ moment per formula unit.

We could achieve excellent fits of the inelastic data of \srni\ after including the interchain couplings. However due to the powder averaging many details of the dispersion is lost thus to unambiguously identify all interchain bonds inelastic neutron scattering on single crystal sample is necessary. The determined coupling constants between spins in the effective model is antiferromagnetic $J_{tri}=1.46(1)$ meV although the individual couplings are ferromagnetic. Using this value we can determine the critical field of the spin flip transition between the hexagonal and the ferromagnetic order. Using mean field theory, the critical field of the field induced transition between the hexagonal and the ferromagnetic structure is:
\begin{align}
	B_C = \frac{6J_{tri}}{M_{Ni}-M_Ir}.
\end{align}
Assuming $g=2$ for both magnetic ions, the spin flip field would be 155 T. This value is in the same order of magnitude as the $B_C=55$ T value found by high field magnetization measurements \cite{Singleton2014}. The difference can be also caused by the unknown $J_{3c}$ exchange between iridium ions.  The experimentally found $0.6\mu_B$ flipped magnetic moment also agrees with the magnetization difference between the ferromagnetic and the hexagonal structure.

\section{Conclusion}

In conclusion, we have investigated \srni\ using inelastic neutron scattering, along with a spin wave analysis. Our INS study reveals spin wave excitations with a giant energy gap of 30 meV at 5 K. More strikingly, these gapped excitations survive up to a high temperature of 200 K, well above $T_N$, thus confirming the quasi-1D nature of the magnetic interaction. Our spin wave analysis has given a good description of the experimental data. Furthermore our fitted values of the anisotropic exchange parameters are in a good agreement with those calculated theoretically using DFT+U+SOC \cite{Gordon}. The presence of the giant spin gap, as compared to the very small spin gap in Sr$_3$ZnIrO$_6$ having only $5d$ magnetic ion (below 1.5 meV with zone boundary energy of 5 meV) reveals that mixed $3d$-$5d$ (or $3d$-$4d$) compounds can generate distinct exchange pathways and can show novel magnetic behavior. Therefore, the present study can foster the research on the magnetic excitations in spin-chain systems to consider such hitherto unrealized factors, and would generate theoretical interest of the development of a more realistic model to understand the complex magnetic behavior of these systems.

\begin{acknowledgments}
We thank Prof. L.C. Chapon, Drs E. Lefrançois, P. McClarty, D. D. Khalyavin, A.D. Hillier, P. Manuel and W. Kockelmann for their involvement. We acknowledge interesting discussion with Profs. S. Lovesey, M.-H. Wangboo and H. Wu. D.T.A. acknowledge financial assistance from CMPC-STFC grant number CMPC-09108. S. T. acknowledges funding from the European Community's Seventh Framework Programme (FP7/2007-2013) under grant agreement n.$^\circ$ 290605  (COFUND: PSI-FELLOW).
\end{acknowledgments}

\appendix

\section{Energy resolution function}
\label{sec:res}
\begin{figure}[!htb]
    \centering
	\includegraphics[width = \linewidth]{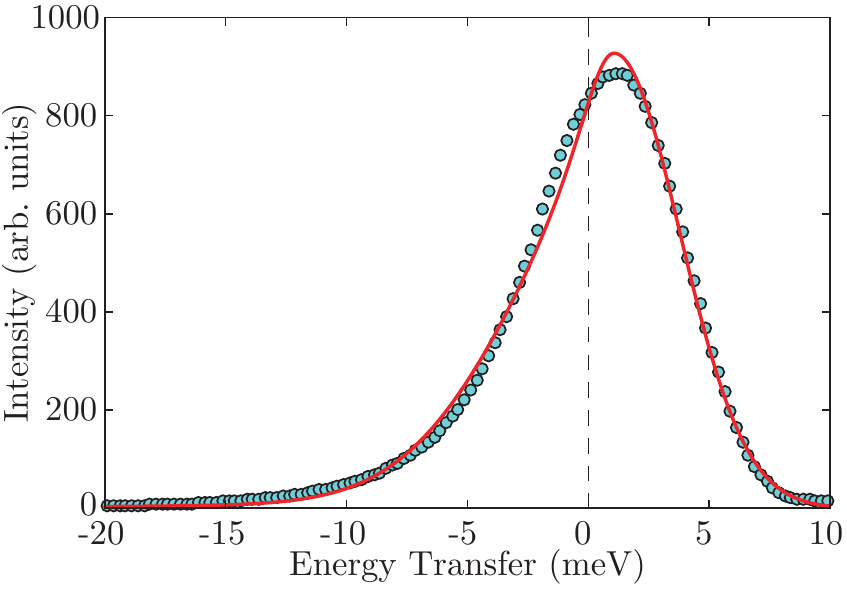}
	\caption{Elastic signal of the sample integrated between 2 \AA$^{-1}$ and 4 \AA$^{-1}$ momentum transfer. The red line denote the resolution function model using a linear combination of different Lorentzian and Gaussian functions on each side.}
	\label{fig:res}
\end{figure}

The energy resolution of a direct time of flight instrument on a neutron spallation source is an asymmetric function with typically a long tail at the low energy side \cite{Ikeda1985}. To model the resolution function, we fitted the inelastic cross section integrated between 2 \AA$^{-1}$ and 4 \AA$^{-1}$ as a function of $E$, see Fig.\ \ref{fig:res}. The peak of the resolution function is not positioned at zero energy transfer due to the asymmetry, but the intensity weighted average should be at zero. We chose a model to fit the peak that is more accurate than a single Gaussian, but simple enough to enable fast convolution of the simulated data. We chose a linear combination of Lorentzian and Gaussian functions with different width on both sides of the peak. The best fit revealed that the high energy side of the peak is purely Gaussian. The fit parameters are shown in Tab.\ \ref{tab:eres}. To account for the resolution change as a function of energy transfer we scaled the width of all components  using the function:
\begin{align}
	w(E) = w(0) E_f^{3/2} = w(0)\cdot (1-E/E_i)^{3/2},
\end{align}
that accounts for the neutron pulse width generated by the chopper system. To keep the integrated intensity of the resolution function constant as a function of $E$, we divide the amplitude with $w(E)$.

\begin{ruledtabular}
\begin{table}[!htb]
	\centering
	\caption{Fit parameters of the energy resolution function of MERLIN at $E_i=80$ meV. $w$ denotes the standard deviation of the Gaussian and the $\gamma$ parameter of the Lorentzian functions, the peak amplitude is normalized to one.}
	\label{tab:eres}
\begin{tabular}{l|l|l|l|l|l|l|l}
$A_1^G$  & $A_2^G$  & $A_1^L$  & $A_2^L$ & $w_1^G$ & $w_2^G$ & $w_1^L$  & $w_2^L$\\
\hline
0.58(2)  & 1        & 0.42(3)  & 0       &3.22(4)  & 1.93(3) & 1.97(10) & -       \\
	\end{tabular}
\end{table}
\end{ruledtabular}

\section{Data treatment and fitting method}
\label{sec:fit}
In order to fit the parameters of a model spin Hamiltonian to inelastic neutron scattering data collected on polycrystalline sample we used the following method. In the first step we removed the phonon scattering from the raw data. This can be done using a few different methods. To remove coherent phonons, we collected the spectrum at high temperature where the magnetic signal is weak and used it as a background for the low temperature data. Before the subtraction we applied an energy dependent scaling to correct for the temperature dependent cross section due to the Bose statistics of phonons and magnons:
\begin{align}
	C(E) = \frac{1-\exp(-E/(k_BT_1))}{1-\exp(-E/(k_BT_2))},
\end{align}
where $T_1$ is the temperature of the phonon data (300 K in our case), while $T_2$ is the temperature of the magnetic data (5 K). After this correction, a weak Q dependent background remained that increased linearly with $Q^2$ and it seemed to originate from incoherent phonon scattering. This survived the previous subtraction probably due to the change in the Debye-Waller factor at high temperature that we did not accounted for. To subtract the incoherent phonons, we averaged the scattering intensity above $Q>8$ \AA$^{-1}$:
\begin{align}
	I^{IC}_j = \frac{1}{N(Q_i>Q_{min},\omega_j)}\displaystyle\sum_{Q_i>Q_{min}}S(Q_i,\omega_j)/Q_i^2,
\end{align}
where $S(Q_i,\omega_j)$ is the measured inelastic scattering intensity on a $(i,j)$ pixel centered at $(Q_i,\omega_j)$ and $N(Q_i>Q_{min},\omega_j)$ is the number of pixels above $Q_{min}$ with a fixed $\omega_j$ value. Afterwards we subtracted the $I_j^{IC}\cdot Q_i^2$ value from each $(i,j)$ pixel. After these corrections the inelastic signal was clean from phonon scattering up to 4 \AA$^{-1}$ see Fig.\ \ref{fig:fit}.

Fitting the complete measured dataset is challenging, due to the computationally intensive powder averaging. To speed up the fitting process, we fitted only a single cut integrated between 2 - 3 \am. The spin wave spectrum was simulated at 5 different $Q$ points evenly distributed in the same range. Due to the weakly dispersive nature of the spectrum as a function of $Q$, the result of the fit is insensitive to the number of $Q$ points averaged. The most common method to calculate the powder average is to use a Monte Carlo technique to average the spectrum over a $|Q|=$const. sphere in reciprocal space. However this method is not ideal for a fitting purpose since the simulated data would contain a noise. We used a deterministic method to generate evenly distributed points on a unit sphere according to Ref. \cite{Hannay2004}. The number of points to average over has to be a Fibonacci number, we chose 987 which gave a reliable average (it gave less than a 1\% error in estimating $\chi_{red}^@$ compared to simulations including more points on the unit sphere).

We used the least squares method to define the goodness of the fit:
\begin{align}
\chi^2 = \displaystyle\sum_i \frac{1}{\sigma_i^2} \left( y_i - y^{LSWT}_i\right)^2,
\end{align}
where $y_i$ are the measured intensities along the cut and $y_i^{SIM}$ are the simulated intensities. Since we do not have reliable data on the upper iridium band, we used the available RIXS data \cite{Lefrancois2015a} to fix the position of the upper iridium band to 95 meV. To constrain the energy of the upper band we summed up all simulated intensity above 65 meV and calculated the center of mass. We added the squared deviation of the upper band to the calculated $\chi^2$ with a large weight which effectively gave a constraint.

To minimize the $\chi^2$ value we used a particle swarm optimization method \cite{Kennedy,Clerc2002}. Unfortunately due to the noisy nature of both the data and the simulation (due to powder averaging) there is no reliable method to extract the standard deviations of the fitted parameters. To compare different simulations with different number of fit parameters we calculated the reduced $\chi^2$:
\begin{align}
\chi_{red}^2 = \frac{\chi^2}{N_{dat}-N_{par}-1},
\end{align}
where $N_{dat}$ is the number of data points along the cut, $N_{par}$ is the number of fitting parameters. For a dataset with reliable error bars a good model fit should give $\chi_{red}^2=1$.


%

\end{document}